\documentclass[showpacs,preprintnumbers]{revtex4}
\usepackage{amsmath}
\usepackage{graphicx}

\input{tcilatex}

\begin{document}

\title{Impact of the transport supercurrent on the density of states in the
weak link}
\author{S.N. Shevchenko}
\email{sshevchenko@ilt.kharkov.ua}
\affiliation{B.Verkin Institute for Low Temperature Physics and Engineering, 47 Lenin
Ave., Kharkov 61103, Ukraine.}
\date{\today}

\begin{abstract}
The impact of the transport supercurrent on the quasiparticle density of
states in the superconducting thin film with the weak link is investigated.
At the weak link the order parameter is locally suppressed due to the order
parameter phase difference $\phi $. This results in the appearance of the
zero-energy states, which are strongly influenced by the supercurrent
especially at $\phi $ close to $\pi $. The subgap density of states,
dependent on both the supercurrent and the phase difference, is shown to
modify the conductance characteristics of the structure.
\end{abstract}

\maketitle

Properties of a mesoscopic superconductor are defined by its local density
of states (LDoS). At the interface LDoS can be probed with the tunneling
spectroscopy \cite{Duke}. Particularly interesting are situations when there
are states (so-called midgap states) within the bulk amplitude value of the
order parameter $\Delta _{0}$, that is at the quasiparticle energy $%
\varepsilon <\Delta _{0}$. This situation is ubiquitous in unconventional
superconductors, where an interface results in the local suppression of the
order parameter (due to its anisotropy) and in the appearance of the
zero-energy states \cite{Hu}. These zero-energy states are responsible for
several interesting phenomena. First, when there is either external magnetic
field or supercurrent injected, these midgap states create the
countercurrent, which counterflows to the diamagnetic current in the former
case, leading to the paramagnetic response \cite{FRS97, Higashitani}, or
counterflows to the injected supercurrent in the latter case \cite{KOSh2}.
Second, when LDoS is probed with the tunneling spectroscopy, they result in
the zero-bias conductance peak \cite{FRS97, Tanaka}.

A weak link between two banks of conventional superconductors is similar to
an interface of an unconventional superconductor in this context of local
suppression of the order parameter \cite{KO} and of the creation of the
midgap states \cite{Furusaki}. In recent years impact of the supercurrent on
the density of states, and in particular, on the midgap states, has
attracted the renewed attention (see e.g. Refs. \cite{Sanchez, Anthore,
Zhang, canadci}). In Refs. \cite{KOSh1} \ we have studied first of the above
mentioned effects in the weak link between current-carrying superconductors.
And in this paper we address the second problem of the appearance of the
midgap states at the weak link between two current-carrying conventional
superconductors. This subject is interesting both for further understanding
of interference effects in mesoscopic superconducting structures and for
possible applications for electronic devices, in which the current can be
strongly controlled either electrodynamically (creating order parameter
phase difference) or thermodynamically (injecting transport current) \cite%
{otherpropos}. In what follows we describe the model system to study the
midgap states at the weak link between two current-carrying conventional
superconductors.

\begin{figure}[tbph]
\includegraphics[width=10cm]{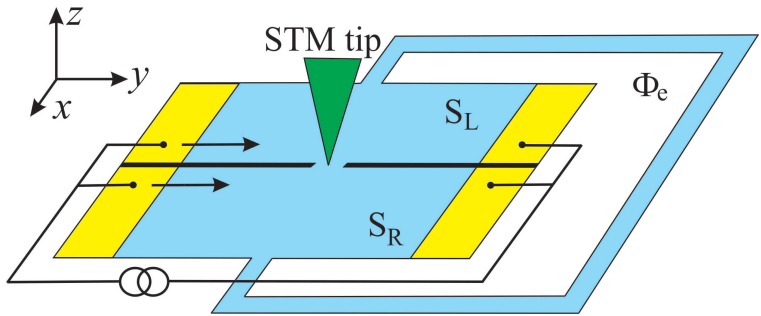}
\caption{(Color online) Scheme for probing LDoS at the point contact between
superconducting banks $S_{L}$ and $S_{R}$ with externally controlled order
parameter phase difference $\protect\phi =2\protect\pi \Phi _{e}/\Phi _{0}$
and externally injected supercurrent in parallel to the boundary between $%
S_{L}$ and $S_{R}$.}
\label{Schema}
\end{figure}

We consider the superconducting film with the impenetrable partition along $%
y $-axis, shown with the thick line in Fig.~\ref{Schema}. 
This partition (which can be \textit{e.g.} a scratch in the film) has a
break, which plays the role of the weak link in the form of a slit between
two superconducting banks $S_{L}$ and $S_{R}$. The size of the weak link is
assumed to be smaller than the coherence length (so-called pinhole model).
The order parameter phase difference $\phi $ at the weak link is created by
the magnetic flux $\Phi _{e}$ which pierces the loop connecting two banks.
The magnetic flux can be created \textit{e.g.}\ via the inductive coupling
to a current-carrying coil. The supercurrent, tangential to the boundary
between the banks $S_{L}$ and $S_{R}$, \ is passed through the contacts, as
in Ref. \cite{canadci}. The LDoS is assumed to be probed by the scanning
tunneling microscope's (STM) tip, placed over the weak link. Another contact
for STM is not shown in Fig. \ref{Schema} and we refer the reader to Ref. 
\cite{canadci} where both the current injection and the STM measurement as
in Fig. \ref{Schema} are discussed in detail.

Thus, we consider a perfect point contact between two clean current-carrying
singlet superconductors. The order parameter phase difference $\phi $ is
assumed to drop at the contact plane at $x=0$. The homogeneous supercurrent
flows in the banks of the contact along the $y$-axis, parallel to the
boundary. The sample is assumed to be smaller than the London penetration
depth so that the externally injected transport supercurrent can indeed be
treated as homogeneous far from the weak link. Such a system can be
quantitatively described by the Eilenberger equations. Taking transport
supercurrent into account leads to the Doppler shift of the energy variable
by $\mathbf{p}_{F}\mathbf{v}_{s}$, where $\mathbf{p}_{F}$\ is the Fermi
momentum and $\mathbf{v}_{s}$\ is the superfluid velocity. The standard
procedure of matching the solutions of the bulk Eilenberger equations at the
boundary gives the Matsubara Green's function $\widehat{G}_{(0)}(\omega )$
at the contact at $x=0$ \cite{KOSh2}. The analytic continuation ($\omega
\rightarrow -i\varepsilon +\gamma $) of the component $G_{(0)}^{11}(\omega
)\equiv g^{(0)}(\omega )$ of $\widehat{G}_{(0)}(\omega )$ gives the retarded
Green's function, which defines LDoS at the boundary:%
\begin{equation}
N(\varepsilon )=\left\langle N(\varepsilon ,\mathbf{p}_{F})\right\rangle
=N_{0}\left\langle \func{Re}g^{(0)}(\varepsilon )\right\rangle ,
\label{N(eps)}
\end{equation}%
\begin{equation}
g^{(0)}(\varepsilon )=g^{(0)}(\omega =-i\varepsilon +\gamma ),
\label{g(eps)}
\end{equation}%
\begin{equation}
g^{(0)}(\omega )=\frac{\widetilde{\omega }(\Omega _{L}+\Omega
_{R})-isgn(v_{x})\Delta _{L}\Delta _{R}\sin \phi }{\Omega _{L}\Omega _{R}+%
\widetilde{\omega }^{2}+\Delta _{L}\Delta _{R}\cos \phi },  \label{g(0)}
\end{equation}%
here $\omega =\pi T(2n+1)$ are Matsubara frequencies, $\gamma $ is the
relaxation rate in the excitation spectrum of the superconductor, $%
N(\varepsilon ,\mathbf{p}_{F})$ is the angle-resolved DoS, $N_{0}$ is the
density of states at the Fermi level, $\left\langle ...\right\rangle $
denotes averaging over the directions of Fermi momentum $\mathbf{p}_{F}$, $%
\Delta _{L,R}$ stands for the order parameter in the left (right) bank, $%
\widetilde{\omega }=\omega +i\mathbf{p}_{F}\mathbf{v}_{s}$, $\Omega _{L,R}=%
\sqrt{\widetilde{\omega }^{2}+\Delta _{L,R}^{2}}$. The poles of the retarded
Green's function $g^{(0)}(\varepsilon )$ define the energy of the interface
bound states. The direction-dependent Doppler shift $\mathbf{p}_{F}\mathbf{v}%
_{s}$ results in the modification of LDoS as it is discussed below.

The method of the tunneling spectroscopy allows to access the LDoS by
measuring the tunneling conductance $G$ of the Superconductor - Insulating
barrier - Normal metal (STM tip) structure. At zero temperature the
dependence of the conductance on the bias voltage $V$ is given by \cite{Duke}%
\begin{equation*}
G(eV)=G_{N}\left\langle D(\mathbf{p}_{F})N(eV,\mathbf{p}_{F})\right\rangle ,
\end{equation*}%
where $G_{N}$ is the conductance in the normal state; $D(\mathbf{p}_{F})$ is
the angle-dependent Superconductor-Insulator-Normal metal barrier
transmission probability. The barrier can be modelled \textit{e.g. }as in
Ref. \cite{FRS97} with the uniform probability within the acceptance cone $%
\left\vert \vartheta \right\vert <\vartheta _{c}$, where $\vartheta $ is the
polar angle and the small value of $\vartheta _{c}$ describes the thick
tunneling barrier. In this paper for simplicity we do not take into account
the angle dependence of the transmission probability, assuming that the
conductance is proportional to the LDoS: $G(eV)\propto N(\varepsilon =eV)$.

In the case of a weak link in the form of a constriction between two
conventional superconductors with $\Delta _{L}=\Delta _{R}=\Delta
_{0}=\Delta _{0}\left( T,\mathbf{v}_{s}\right) $ \cite{KOSh1} Eq. (\ref{g(0)}%
) can be rewritten:

\begin{equation}
g^{(0)}(\omega )=\frac{\widetilde{\omega }\Omega -i\frac{1}{2}%
sgn(v_{x})\Delta _{0}^{2}\sin \phi }{\widetilde{\omega }^{2}+\Delta _{\phi
}^{2}},  \label{g_s-wave}
\end{equation}%
where $\Delta _{\phi }=\Delta _{0}\left\vert \cos (\phi /2)\right\vert $ is
the locally suppressed order parameter (so-called proximity gap) \cite{KO}.
The retarded Green's function, determined with Eqs. (\ref{g(eps)}) and (\ref%
{g_s-wave}), gives for the energy of the Andreev bound states: $\varepsilon
_{A}=\pm \Delta _{\phi }-\mathbf{p}_{F}\mathbf{v}_{s}$. Thus, there are the
zero energy states, which are characterized by the values $\phi $ and $v_{s}$%
. The observation of the conductance characteristics of the system can be
proposed as a test of the interface-induced transport-current-dependent
quasiparticle states \cite{FRS97, Furusaki, KOSh1}.

For the sake of generality we also write down below the Green's function for
the current-carrying $d$-wave superconductor. This can be easily done in the
case of the specular reflection at the boundary, when the boundary between
the current-carrying $d$-wave superconductor and the insulator can be
modelled as the contact between two superconductors with the order
parameters given by $\Delta _{L}=\Delta (\vartheta )$ and $\Delta
_{R}=\Delta (-\vartheta )\equiv \overline{\Delta }$ and with $\phi =0$. Then
from Eq. (3) we have the following expression, which together with Eq. (\ref%
{N(eps)}) describes the LDoS in the current-carrying $d$-wave film as 
\textit{e.g.} in Refs. \cite{FRS97, canadci}:%
\begin{equation}
g^{(0)}(\omega )=\frac{\widetilde{\omega }\left( \Omega +\overline{\Omega }%
\right) }{\Omega \cdot \overline{\Omega }+\widetilde{\omega }^{2}+\Delta
\cdot \overline{\Delta }},  \label{g_for_d-film}
\end{equation}%
where $\Omega =\sqrt{\widetilde{\omega }^{2}+\Delta ^{2}}$ and $\overline{%
\Omega }=\sqrt{\widetilde{\omega }^{2}+\overline{\Delta }^{2}}$. This
expression is valid for any relative angle $\chi $ between the $a$-axis and
the normal to the boundary; in particular, at $\chi =0$ we have: $%
g^{(0)}(\omega )=\widetilde{\omega }/\Omega $, and at $\chi =\pi /4$ we
have: $g^{(0)}(\omega )=\Omega /\widetilde{\omega }$. Analogously Eq. (3)
can be used to describe the LDoS at the boundary between two
current-carrying $d$-wave superconductors as in Fig. 1 (see also in Ref. 
\cite{KOSh2}).

In what follows we consider the LDoS at the point contact between
current-carrying conventional superconductors in details first analytically
at $\gamma =0$ and then numerically for $\gamma \neq 0$.

In the absence of impurities, that is in the limit $\gamma \rightarrow +0$,
we obtain the LDoS depending on the energy $\varepsilon $ and on the
parameters $Q=p_{F}v_{s}$ and $\phi $:%
\begin{equation}
\frac{N(\varepsilon ,Q,\phi )}{N_{0}}=\left\langle \frac{\left\vert 
\widetilde{\varepsilon }\right\vert \sqrt{\widetilde{\varepsilon }%
^{2}-\Delta _{0}^{2}}}{\widetilde{\varepsilon }^{2}-\Delta _{\phi }^{2}}%
\cdot \theta \left( \left\vert \widetilde{\varepsilon }\right\vert -\Delta
_{0}\right) +\frac{\pi }{2}\Delta _{S}\cdot \delta \left( \left\vert 
\widetilde{\varepsilon }\right\vert -\Delta _{\phi }\right) \right\rangle ,
\label{N_for_SoS}
\end{equation}%
where $\theta (...)$ and $\delta (...)$ are the theta- and delta- functions; 
$\widetilde{\varepsilon }=\varepsilon -\mathbf{p}_{F}\mathbf{v}_{s}$; $%
\Delta _{S}\equiv \Delta _{0}\left\vert \sin \left( \phi /2\right)
\right\vert $. This after integration results in:%
\begin{eqnarray}
\frac{N(\varepsilon ,Q,\phi )}{N_{0}} &=&\sum\limits_{\pm }(\pm 1)sgn\left(
\varepsilon _{\pm }\right) \theta \left( \varepsilon _{\pm }^{2}-\Delta
_{0}^{2}\right) \frac{\sqrt{\varepsilon _{\pm }^{2}-\Delta _{0}^{2}}}{2Q}+
\label{LDoS_for_SoS} \\
&&+\frac{\Delta _{S}}{2Q}\sum\limits_{\pm }\left[ \left( \mp 1\right)
sgn\left( \varepsilon _{\pm }\right) \theta \left( \varepsilon _{\pm
}^{2}-\Delta _{0}^{2}\right) \arctan \frac{\sqrt{\varepsilon _{\pm
}^{2}-\Delta _{0}^{2}}}{\Delta _{S}}+\frac{\pi }{2}\theta \left(
Q-\left\vert \varepsilon \pm \Delta _{\phi }\right\vert \right) \right] . 
\notag
\end{eqnarray}%
Here $\varepsilon _{\pm }\equiv \varepsilon \pm Q$; the first term describes
the LDoS related to the continuum states, which coincides with the LDoS in
the current-carrying homogeneous thin film \cite{Fulde}, and the second term
gives the contribution of the bound states.

Results of the numerical calculation at non-zero relaxation rate $\gamma $
are presented in Fig.~2. 
\begin{figure}[tbph]
\includegraphics[width=10cm]{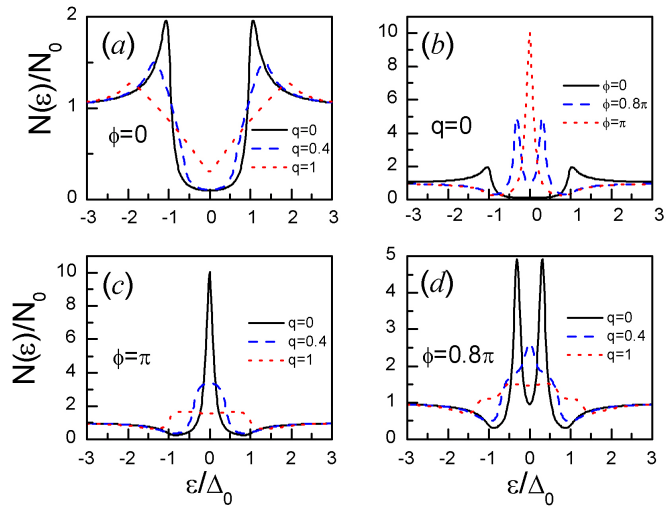}
\caption{(Color online) LDoS for the point contact between two
current-carrying conventional superconductors for different values of the
phase difference $\protect\phi $ and the dimensionless supercurrent velocity 
$q=p_{F}v_{s}/\Delta _{0}$ at zero temperature $T=0$ and non-zero relaxation
rate $\protect\gamma /\Delta _{0}=0.1$.}
\end{figure}
First, LDoS in the homogeneous current state (that is at $\phi =0$) is shown
in Fig. 2(a) to experience the suppression of the peak at $\varepsilon
=\Delta _{0}$ and gradual appearance of the midgap states with increasing
the transport supercurrent \cite{Fulde}. Then LDoS in the point contact
without the supercurrent ($q=0$) is plotted in Fig.2(b). With increasing the
phase difference $\phi $ from $0$ to $\pi $ the peak is shifted from $%
\varepsilon =\Delta _{\phi =0}=\Delta _{0}$ to $\varepsilon =\Delta _{\phi
=\pi }=0$. Note that the peak at $\varepsilon =0$ would appear as the
so-called zero-bias conductance peak (ZBCP) in the STM measurements (see
also in Ref. \cite{Amin} about the appearance of ZBCP in point-contact
Josephson junctions with ac biasing voltage). In Fig. 2(c) it is shown that
the supercurrent results in the suppression and widening of this ZBCP
similar to the ZBCP suppression in \textit{d}-wave superconductor \cite%
{canadci}. Note that in concrete realizations the details of the
modification of the ZBCP (widening, splitting) depend on several parameters,
such as barrier anisotropy, surface roughness \cite{FRS97, canadci, Aubin}.
And finally in Fig. 2(d) we study the situation when the order parameter at
the contact is significantly suppressed due to the phase difference at $\phi 
$ close to $\pi $. In the absence of the supercurrent there are two midgap
peaks (see also in Fig. 2(b)). With the gradual increase of the supercurrent
these two peaks are shifted and modified so that first three-humped and then
four-humped midgap structures appear, which are shown respectively for $%
q=0.4 $ and for $q=1$. Similar humped structure of LDoS in homogeneous
current-carrying superconductor (\textit{i.e.} at $\phi =0$) was studied in
Ref. \cite{Zhang}. In the homogeneous situation, studied in Ref. \cite{Zhang}%
, the particularly interesting zero-energy states (which appear as ZBCP)
take place when the supercurrent is strong enough to depair electrons but
smaller than the thermodynamic critical current, that is in rather narrow
region: $\Delta _{0}<p_{F}v_{s}<1.03\Delta _{0}$. We emphasize that in our
case when the order parameter is locally suppressed (controlled) by the
phase difference, the ZBCP appears at any value of the supercurrent at the
phase difference defined by the relation $\Delta _{\phi }=\Delta _{0}\cos
(\phi /2)<p_{F}v_{s}$ (see Figs. 2(b) and (d)).

In conclusion, LDoS at the weak link between current-carrying
superconductors have been studied. We propose this LDoS to be visualized
with the STM. Two controlling mechanisms - external magnetic flux and
injected supercurrent - allow to vary LDoS, and correspondingly the
characteristics of the structure in different ways in the wide range of the
parameters $\phi $ and $v_{s}$.

The author is grateful to A.N. Omelyanchouk, Yu.A. Kolesnichenko, and M.
Grajcar for helpful discussions. This work was partly supported by grant of
President of Ukraine (No. GP/P11/13).

\end{document}